\documentstyle{article}
\begin{document}
\title{Noised based Cipher system.} \author{Arindam Mitra
\\V.I.P Enclave, M - 403, Calcutta- 700059.  India.\\}
\date{}
\maketitle
\begin{abstract}\bf
A computationally secure noised based cipher system is proposed.
The advantage of this cipher system is that it 
operates above noise level. Therefore computationally 
secure communication can be done 
when error correction code fails. Another feature of this system is that
 minimum number of exhaustive key 
search can be made fixed.
 
 \end{abstract} 
\newpage
\section*{}
All the popular classical cipher systems [1-2] are computationally secure.
As computational security is not 
 mathematically proven, it does not always satisfy all. Nevertheless
computational secure systems are used when 'cover time'  is not long.
But these cryptosystem can not be used if the  classical 
communication system becomes highly corrupted  which classical
error correcting code cannot correct. 
We shall see, the solution of the above problem lies in the noise itself.
Manipulating noise a simple cipher system is presented.
The cipher system works on  the same  coding and decoding technique developed for quantum cryptography [3].

\paragraph*{}In our coding decoding procedure, 
two different sequences of  states represent two 
bit values. The information regarding the two  sequences  
are shared between the legitimate users. 
Sender  repeatedly and randomly uses the two sequences  to
produce an arbitrarily long string of bits (i.e. the key)
 and receiver recovers the bit values from the shared information.

\paragraph*{}
Firstly we shall describe the error introduction strategy 
in a simple classical system.
The basic idea behind  this protocol is that a 
sequence of one type of element 
will be corrupted in two different ways to  represent two bit values. 
This is equivalent
to choosing two sequences for two bit values.
Suppose the  sequence  is \\
$ \it S =\left\{1, \,\, 1, \,\,1, \,\,1, \,\,1, \,\,1, \,\,1, \,\,1, \,\,1, \,\,1, \,\,
1, \,\,1,  ........\right\}$. This sequence can be represented by  
two different sub sequences (denoted by $\alpha$ and $\beta$) of 
elements {\it 1}.
$ \it S = \left\{\alpha, \,\,\beta, \,\,\beta, \,\,\alpha, \,\,\beta, \,\,
\alpha, \,\,\beta, \,\,\beta, \,\,\alpha, \,\,\alpha, \,\, \beta, \,\,\alpha,.......\right\}$,
where probability of $\alpha$ position is same with the probability of 
$\beta$ position ($p_{\alpha}= p_{\beta}$) in the sequence {\it S}. 
Information regarding 
the sub sequences $\alpha$ and $\beta$ are shared between sender and receiver.
 Apart 
from  environmental noise these sub sequences of
$\alpha$ and $\beta$ can be further corrupted by two  error-introducing systems X and Y.
Considering natural error, two different probabilistic error  $p_{x}$ and 
$p_{y}$ can be fixed by X and Y. 
For bit {\bf 0}, sender  introduces error $p_{x}$ and $p_{y}$ in the 
sub sequences 
$\alpha$ and $\beta$ respectively. Now sequence {\it S} can be 
termed as probabilistic sequence, $S^{p}$, where  each element is denoted 
by its error probability. 
$S_{0}^{p} =\left\{ p_{x}, \,\, p_{y}, \,\, p_{y},  \,\,p_{x},  \,
\,p_{y},  \,\,p_{x},  \,\,p_{y},  \,\,p_{y},  \,\,p_{x},  \,\,p_{x}, \,\,p_{y}, \,\,p_{x},  
....\right\}$ Similarly, for bit {\bf 1}, error $p_{x}$ and $p_{y}$  
will be introduced in the sub sequences $\beta$
and $\alpha$ respectively, then $S^{p}$ will be 
$ S_{1}^{p} =\left\{ p_{y}, \,\, p_{x},  \,\,p_{x},  \,\,p_{y},  
\,\,p_{x},  \,\,p_{y},  \,\,p_{x},  \,\,p_{x},  \,\,p_{y}, \,\,p_{y}, \,\,p_{x}, \,\,p_{y},
.....\right\}$.

The  key will be:
\begin{eqnarray} K = \left(\begin{array}{ccccccccccccc}
p_{x} & p_{y} & p_{y} & p_{x} & p_{y} & p_{x} & p_{y} &p_{y} & p_{x} & p_{x} & p_{y} & p_{x}  & ....\\
p_{y} & p_{x} & p_{x} & p_{y} & p_{x} & p_{y} & p_{x} & p_{x} & p_{y} & p_{y} & p_{x} & p_{y} & ....\\
p_{y} & p_{x} & p_{x} & p_{y} & p_{x} & p_{y} & p_{x} & p_{x} & p_{y} & p_{y} & p_{x} & p_{y} & ....\\
p_{x} & p_{y} & p_{y} & p_{x} & p_{y} & p_{x} & p_{y} & p_{y} & p_{x} & p_{x} & p_{y} & p_{x} & ....\\
p_{x} & p_{y} & p_{y} & p_{x} & p_{y} & p_{x} & p_{y} & p_{y} & p_{x} & p_{x} & p_{y} & p_{x} & ....\\
p_{y} & p_{x} & p_{x} & p_{y} & p_{x} & p_{y} & p_{x} & p_{x} & p_{y} & p_{y} & p_{x} & p_{y} & ....\\
.     & .     & .     &   .   &   .   &   .   &  .    &  .    &   .   &   .   &   .   &  .    & ....\\ 
.     & .     & .     &   .   &   .   &   .   &  .    &  .    &   .   &   .   &   .   &  .    & ....\\ 
.     & .     & .     &   .   &   .   &   .   &  .    &  .    &   .   &   .   &   .   &  .    & .... 
\end{array}\right) \equiv \left(\begin{array}{c} 
\bf 0 \\ \bf 1 \\ \bf 1 \\ \bf 0 \\ \bf 0 \\ \bf 1 \\. \\. \\. 
\end{array}\right)\nonumber\end{eqnarray}

\noindent
All  rows (columns also) are having same error $p_{z}$, where $p_{z} 
=1/2 (p_{x} + p_{y}) $. The receiver can easily distinguish the bit values
 of the corrupted sequences as he knows 
the sub sequences  $\alpha$ and $\beta$ provided bits are statistically distinguishable.
But code breaker has 
to go through the exhaustive search as he/she does not know the sub sequences  
$\alpha$ and $\beta$. By exhaustive search code breaker
can find the two 
error levels $p_{x}$ and $p_{y}$ for two different 
sub sequences of $\alpha$ and $\beta$ (statistical problem of
exact search is ignored).
Once code breaker knows the position of  $\alpha$ and $\beta$, 
he/she can easily break the code.
Of course
 code breaker can pursue the exhaustive
search with marginal number of bits. This will effectively reduce the
number of search. Using the concept of pseudo-bit (fake bit), 
we shall now discuss how to
defeat the minimal exhaustive search.

\paragraph*{}
Here {\it S} can be represented as a composition of three sub sequence of $\alpha$, $\beta$ and $\gamma$. \\
 So $ S= \left\{\gamma, \,\,\gamma, \,\,\gamma, \,\,\gamma, \,\,\gamma, \,\,\alpha, \,\,\beta, 
\,\,\gamma, \,\,\gamma, \,\,\gamma, \,\,\gamma, \,\,\gamma, \,\,\gamma, \,\,\beta, \,\,\alpha, \,\,\gamma, ....\right\}$,
where $p_{\alpha} = p_{\beta}$, but less than $p_{\gamma}$, the probability
of  $\gamma$ position in the sequence of {\it S}. 
Similarly  sender and receiver share the information of the 
sub sequences $\alpha$, 
$\beta$ and $\gamma$. Apart from the devices X and
Y, we need another error introducing device Y which can produce error $p_{z}$.
Now error $p_{z}$, $p_{x}$ and $p_{y}$ will be introduced in the sub sequences 
 $\gamma$, $\alpha$ and $\beta$ respectively to represent bit {\bf 0}
and error $p_{z}$, $p_{y}$ and $p_{x}$ will be introduced in the 
sub sequences  $\gamma$, $\beta$ and $\alpha$ respectively to represent bit {\bf 1}, where
$p_{z} =1/2 (p_{x} + p_{y}) $. 
Then the extended key is: 
\begin{eqnarray} K^{E}= \left(\begin{array}{ccccccccccccccccc}
p_{z} & p_{z} & p_{z} & p_{z} & p_{z} & p_{x} & p_{y} & p_{z} & p_{z} &  p_{z} & p_{z} & p_{z}  & p_{y} & p_{x} & p_{z} & p_{z} &  ....\\
p_{z} & p_{z} & p_{z} & p_{z} & p_{z} & p_{y} & p_{x} & p_{z} & p_{z} &  p_{z} & p_{z} & p_{z}  & p_{x} & p_{y} & p_{z} & p_{z} &  ....\\
p_{z} & p_{z} & p_{z} & p_{z} & p_{z} & p_{y} & p_{x} & p_{z} & p_{z} &  p_{z} & p_{z} & p_{z}  & p_{x} & p_{y} & p_{z} & p_{z} &  ....\\
p_{z} & p_{z} & p_{z} & p_{z} & p_{z} & p_{x} & p_{y} & p_{z} & p_{z} &  p_{z} & p_{z} & p_{z}  & p_{y} & p_{x} & p_{z} & p_{z} &  ....\\
p_{z} & p_{z} & p_{z} & p_{z} & p_{z} & p_{x} & p_{y} & p_{z} & p_{z} &  p_{z} & p_{z} & p_{z}  & p_{y} & p_{x} & p_{z} & p_{z} &  ....\\
p_{z} & p_{z} & p_{z} & p_{z} & p_{z} & p_{y} & p_{x} & p_{z} & p_{z} &  p_{z} & p_{z} & p_{z}  & p_{x} & p_{y} & p_{z} & p_{z} &  ....\\
. & . & . & . & . & . & . & . & . &  . & . & .  & . & . & . & . &  ....\\ 
. & . & . & . & . & . & . & . & . &  . & . & .  & . & . & . & . &  ....\\ 
. & . & . & . & . & . & . & . & . &  . & . & .  & . & . & . & . &  .... 
\end{array}\right) \equiv \left(\begin{array}{c} 
\bf 0 \\ \bf 1 \\ \bf 1 \\ \bf 0 \\ \bf 0 \\ \bf 1 \\. \\. \\. \end{array} \right)\nonumber\end{eqnarray}

\noindent
Again the probability of error in  columns (also in rows) is same 
which is $p_{z}$. 
  Due to the presence of pseudo-elements (position of $\gamma $),  
code breaker's marginal statistics   will have to be increased.
Increasing the number of pseudo-elements, we can force code breaker
to pursue  more lengthy exhaustive search.
 As  $\gamma$ sub sequence is a pseudo sequence, receiver can only measure
on  $\alpha$ and $\beta$
sub sequences to reveal the bit values.

\paragraph*{}
It is trivial to  mention that
single sequence $S$  can be a  binary sequence. Then a 
sequence of binary sources
will be corrupted in two different ways to produce two values.
 The error introduction strategy is also 
applicable for quantum source. But measured error 
in quantum states
is  not independent to measurements - it is a function  measurement.  
 
\paragraph*{}
Let us assume $p_{z} = 1/2 $, $p_{x} = 1/4 $ 
and $p_{y} = 3/4$  and the numbers of $\gamma$, $\alpha$ and $\beta$ are
$10^{4}$, $10^{2}$ and $10^{2}$ respectively i.e. $p_{\gamma} : p_{\alpha} 
: p_{\beta} = 100:1:1$. 
So code breaker's  minimum statistics for exhaustive 
key search should  
be greater than 100. 
Therefore code breaker has to undertake more than $2^{100}$ operations. 
 The practical advantage of this system
is that secure message can be directly transmitted.
As noisy apparatus is needed,  secure 
communications might be economical. \\

\noindent
{\bf Note added:}
Using the concept of "pseudo-bit" and without using noise,
a simple unbreakable code
has recently been proposed [4].

\newpage

 \end{document}